\documentclass[a4paper,11pt,leqno,twoside]{article}

\usepackage{amsmath, amsthm, amssymb,bm}
\numberwithin{equation}{section}

\newcommand{\bC}{\mathbb{C}}

\newcommand{\bQ}{\mathbb{Q}}

\newcommand{\mf}[1]{\mathfrak{#1}}
\newcommand{\mr}[1]{\mathrm{#1}}
\newcommand{\mcal}[1]{\mathcal{#1}}

\def\tr{\mathrm{tr}}

\def\({ \left( }
\def\){ \right)}

\theoremstyle{plain}
\newtheorem{thm}{Theorem}[section]
\newtheorem{prop}[thm]{Proposition}
\newtheorem{lem}[thm]{Lemma}

\theoremstyle{definition}

\newtheorem{example}{Example}[section]

\theoremstyle{conjecture}
\newtheorem{conj}[thm]{Conjecture}
\theoremstyle{problem}

\title{\bfseries On some properties of orthogonal Weingarten functions}
\author{Beno\^{i}t Collins and Sho Matsumoto}
\date{\today}

\begin{document}

\maketitle

\begin{abstract}
We give a Fourier-type formula for computing the orthogonal Weingarten formula. 
The Weingarten calculus was introduced 
as a systematic method to compute 
integrals of polynomials with respect to Haar measure
over classical groups.
Although a Fourier-type formula was known
 in the unitary case, the orthogonal counterpart was not known.
It  relies on the
Jack polynomial generalization of both 
Schur and zonal polynomials.
This formula substantially reduces the complexity involved
in the computation of Weingarten formulas.
We also describe a few more new properties of the Weingarten formula,
state a conjecture and give a table of values.
\end{abstract}

\section{Introduction}

The terminology of Weingarten calculus was introduced in \cite{Collins}, and
 designates the formula involved in the computation of the integral of
polynomial functions over compact matrix groups with respect to 
their Haar measure. 
Weingarten calculus is an analogy of Wick calculus, which tells 
that the joint moments of random vectors of a real Gaussian space $V$
can be computed by the formula
$$E(v_1\ldots v_{2n})=\sum_{\mf{m}\in \mcal{M}(2n)} E(v_{\mf{m}(1)}v_{\mf{m}(2)})\ldots
E(v_{\mf{m}(2k-1)}v_{\mf{m}(2k)}),$$
where $v_i$ are gaussian vectors of $V$, $E$ is the expectation, and
$\mcal{M}(2n)$ is the set of pair partitions of $\{1,\ldots ,2n\}$ (see section
\ref{sec:reminders} for the notations and definitions).
This formula covers the computation of all moments,
as odd moments are zero due to the $1\to -1$ symmetry.

The question of systematically computing the Haar measure on polynomials 
originally emanates from theoretical physics. One of the most notable attempts is due 
to Weingarten in \cite{Weingarten}, and this is why this calculus bears his name. 
In the paper \cite{Collins}, only the case of integration over the unitary 
group is being considered, 
with a restriction on the bound of the polynomial.
This restriction is relieved later in the paper \cite{CollinsSniady}, where 
an orthogonal version of the Weingarten was developed. 

However, whereas in \cite{Collins}, an explicit Fourier type formula for the unitary Weingarten
function is given as a sum on the irreducible summands of the $n^{th}$  tensor power of the
fundamental representation of the unitary group, 
we were not able to do so in a satisfactory way for the orthogonal group case, although
the canonical description of the irreducible summands of the $n^{th}$ tensor power of
the funamental representation of the orthogonal group is very similar to the unitary group
case. 

Our present paper addresses this question.
The solution comes through the observation that the Schur polynomials involved in
the Fourier-type formula for the Weingarten formula in the unitary case are actually a special
case of so called Jack polynomials. 

The organization of this paper is as follows:
in section \ref{sec:reminders}, we recall the Weingarten formula in the orthogonal case. 
For the sake of completeness we give a complete proof, adapted from \cite{BanicaCollins}.
This proof is much simpler than the original one.
In section \ref{sec:new-expression}, we state our main result and then prove it in
section \ref{sec:proof}.
In section \ref{sec:truncated}, we consider as an application 
the example of a truncated orthogonal random orthogonal matrix. 
In section \ref{sec:full-cycle}, we study the specific case of 
the orthogonal Weingarten function evaluated pairings that form together a full
cycle. This leads us to some new properties of the orthogonal Weingarten functions,
and to a conjecture.
We finish with section  \ref{sec:tableofvalues}, where we give a table of values
of the Weingarten function.

We would like to point out that the case of the symplectic group can be treated in a similar
fashion. We intend to deal with that case in a forthcoming paper.

\section{Weingarten convolution formula for orthogonal and unitary groups}
\label{sec:reminders}

In this section we recall the Weingarten formula for 
orthogonal groups. 
We finish this section by recalling the unitary formula. 
Chronologically, the unitary formula was established earlier, 
but our approach makes it simpler to consider the orthogonal case first.

Let $\mcal{M}(2n)$ be the set of all pair partitions on $\{1,2,\dots,2n\}$.
Each pair partition $\mf{m}$ in $\mcal{M}(2n)$ is uniquely expressed by the form
\begin{equation} \label{eq:ExpressionPairPartition}
\left\{ \{\mf{m}(1),\mf{m}(2) \}, \{\mf{m}(3),\mf{m}(4) \}, \dots, \{\mf{m}(2n-1),\mf{m}(2n) \}
\right\}
\end{equation}
with $\mf{m}(2i-1) < \mf{m}(2i)$ for $1 \le i \le n$ and with
$\mf{m}(1) <\mf{m}(3) < \cdots < \mf{m}(2n-1)$.

Given two pair partitions $\mf{m},\mf{n} \in \mcal{M}(2n)$, we define 
the graph $\Gamma(\mf{m},\mf{n})$ as follows.
The vertex set of $\Gamma(\mf{m},\mf{n})$ is $\{1,2,\dots,2n \}$
and the edge set consists of $\{\mf{m}(2i-1), \mf{m}(2i)\}$ and 
$\{\mf{n}(2i-1), \mf{n}(2i)\}$ with $1 \le i \le n$.
The {\it Gram matrix} $G^{O(d)}_n=(G^{O(d)}(\mf{m},\mf{n}))_{\mf{m},\mf{n} \in \mcal{M}(2n)}$ 
is given by
$$
G^{O(d)}(\mf{m},\mf{n})= d^{\mr{loop}(\mf{m},\mf{n})},
$$
where $\mr{loop}(\mf{m},\mf{n})$ is the number of loops in 
the graph $\Gamma(\mf{m},\mf{n})$, i.e., 
the number of connected components of $\Gamma(\mf{m},\mf{n})$.

Let $\mr{Wg}^{O(d)}_n=(\mr{Wg}^{O(d)}(\mf{m},\mf{n}))_{\mf{m},\mf{n} \in \mcal{M}(2n)}$
be the pseudo-inverse matrix of $G_n^{O(d)}$.
We recall that for any real symmetric
matrix $A$ there exists a unique matrix $B$ satisfying
$ABA=A$ and $BAB=B$. If $A$ is invertible, then $B=A^{-1}$.
The matrix $B$ is called the {\em pseudo-inverse} of $A$.
We call $\mr{Wg}^{O(d)}_n$ the {\it Weingarten matrix} for the orthogonal group.
The following Theorem was proved in \cite{CollinsSniady}:

\begin{thm}[\cite{CollinsSniady}] \label{Thm:CS}
Given indices $i_1,\dots,i_{2n}$ and $j_1,\dots,j_{2n}$ in 
$\{1,2,\dots,d\}$,
\begin{equation}
\int_{g \in O(d)} g_{i_1 j_1} \cdots g_{i_{2n} j_{2n}} d g =
\sum_{\mf{m}, \mf{n} \in \mcal{M}(2n)} 
\mr{Wg}^{O(d)}(\mf{m},\mf{n})
\prod_{k=1}^n \delta_{i_{\mf{m}(2k-1)}, i_{\mf{m}(2k)}} \delta_{j_{\mf{n}(2k-1)}, j_{\mf{n}(2k)}}. 
\end{equation}
\end{thm}

Observe also that in the odd case, we always have
$\int_{g \in O(d)} g_{i_1 j_1} \cdots g_{i_{2n+1} j_{2n+1}} d g =0$
as the Haar measure is invariant under the transformation $1\to -1$.
Let us illustrate this Theorem by a few examples:

\begin{example}
\begin{itemize}
\item
Integrands of type $g_{i_1 j_1} \cdots g_{i_{2n} j_{2n}} $
which don't have pairings
integrate to zero. For example,
$\int_{g \in O(d)} g_{11}^3g_{12} d g =\int_{g \in O(d)} g_{11}^3g_{22} d g =0$.
\item
The integral below admits only one pairing ($1\to 1$ and $2\to 2$).
Therefore,
$$\int_{g \in O(d)} g_{11}^2g_{22}^2 d g =\mr{Wg}^{O(d)}( \{ \{1,\, 2\},\{3,\,4\}\},  \{\{1,\, 2\},\{3,\,4\}\}).$$
A direct computation shows that this coefficient of $\mr{Wg}^{O(d)}_4$ is 
$\frac{d+1}{d(d+2)(d-1)}$
(alternatively, see Section \ref{sec:tableofvalues}).
Therefore
$$\int_{g \in O(d)} g_{11}^2g_{22}^2 d g =\frac{d+1}{d(d+2)(d-1)}$$
for any $d\geq 2$.
\item
For the integral
$\int_{g \in O(d)} g_{11}^4 d g$, all pairings are admissible (because all the indexes 
are the same, equal to $1$). Therefore the integral is the sum of all entries of
$\mr{Wg}^{O(d)}_4$.
A direct computation shows that the answer is
$\frac{3}{d(d+2)}$ for all $d$. Observe that this is indeed the second moment of 
a hyperspherical distribution.
\end{itemize}
\end{example}

Consider the vector space where $O(d)$ acts, namely
$$V={\mathbb C}^d$$
and denote by $e_1,\ldots ,e_d$ its standard basis. 
We need to introduce an embedding of $\mcal{M}(2n)$
in $V^{\otimes 2n}$, which we shall denote by $\rho$.
This embedding is defined by the formula
$$\rho (\mf{m})=\sum_{j_1\ldots j_{2n}}
\begin{pmatrix} \mf{m}\cr j_1\ldots j_{2n} \end{pmatrix}e_{j_1}\otimes\ldots\otimes e_{j_{2n}},$$
where the symbol $\(\begin{smallmatrix} \mf{m}\cr j_1\ldots j_{2n} \end{smallmatrix}\)$ here 
is considered as 
a generalized Kronecker delta symbol: it is
$1$ if all strings of $\mf{m}$ join pairs of equal indices, and is $0$ if not. 
\begin{example}
If $\mf{m}= `` \cap " \in \mcal{M}(2)$, then
$$\rho(\mf{m})=
\sum_{j_1, j_2}\begin{pmatrix} \cap \cr j_1, j_2\end{pmatrix}
e_{j_1}\otimes e_{j_2}=\sum_{j_1, j_2}\delta_{j_1,j_2}e_{j_1}\otimes e_{j_2}
=\sum_{j=1}^de_{j}\otimes e_{j}.$$
\end{example}
Although we won't need this fact, 
let us remark that  $\rho$ is an action in a Brauer algebra sense. 
We will also recover in the course of the proof of
Theorem  \ref{thm:orthogonalWg},
the fact that the family
 $\{\rho (\mf{m}),\mf{m}\in \mcal{M}(2n)\}$ is an independent family 
 provided that $d\geq n$.

We denote by $Fix_{O(d)}V^{\otimes n}$ the vector subspace of 
$V^{\otimes n}$ of invariant elements
under the action of $O(d)$, namely,  elements $x$  that satisfy for all $g\in O(d)$,
$g^{\otimes n}\cdot x=x$.
The main ingredient for the proof of Theorem \ref{Thm:CS}
is the following classical result, 
sometimes known as {\it orthogonal Schur-Weyl duality} formula.
We refer to \cite{GoodmanWallach} for a proof and details.

\begin{prop} [see, e.g. \cite{GoodmanWallach}]
\label{prop:orthogonal-schur-weyl}
In the odd case, we have
$Fix_{O(d)}(V^{\otimes 2n+1})=\{0\}$.
In the even case, we have
$$Fix_{O(d)}(V^{\otimes 2n})=span \{\rho (\mf{m}), \mf{m}\in \mcal{M}(2n)\}.$$
\end{prop}
The above proposition is the main ingredient to the following lemma:

\begin{lem}
\label{lem-intermediate}
The element
$P=\int_{g\in O(d)}g^{\otimes n}dg\in \mathbb{M}_d(\mathbb{C})^{\otimes n}$ is the orthogonal
projection. Its image is exactly 
$span \{\rho(\mf{m}), \mf{m}\in \mcal{M}(2n)\}$. 
\end{lem}

\begin{proof}
First, observe that due to the invariance of the Haar measure,
$$P^2=\int_{g_1,g_2\in O(d)}g_1^{\otimes n}g_2^{\otimes n}dg_1dg_2
=\int_{g_1,g_2\in O(d)}(g_1g_2)^{\otimes n}dg_1dg_2
=\int_{g\in O(d)}g^{\otimes n}dg=P$$
Similarly,
$$P^*=\int_{g^*\in O(d)}g^{\otimes n}dg
=\int_{g^{-1}\in O(d)}g^{\otimes n}dg=\int_{g\in O(d)}g^{\otimes n}dg=P.$$
So, $P$ is an orthogonal projection.
If $n$ is odd, the $1\to -1$ symmetry tells that the above integral is $0$,
therefore $P$ is the zero projection.
In the even case, 
by definition and due to proposition
\ref{prop:orthogonal-schur-weyl}, an element $x\in V^{\otimes 2n}$ 
satisfies 
$g^{\otimes 2n}\cdot x=x$ iff
$x\in span \{\rho(\mathfrak{m}), \mathfrak{m} \in \mcal{M}(2n)\}$. 
This proves that the image of $P$ is as claimed in the statement of the lemma.
\end{proof}

The following lemma could stand as an euclidean geometry course exercise, 
and explains how to compute the coordinates of a projection onto the span of
a set.

\begin{prop}
\label{prop:projection}
In an euclidean (or Hermitian) space
with scalar (resp. Hermitian) product $<\cdot , \cdot >$, 
let $v$ be an element and $P$ be an orthogonal projection, and
$(v_1,\ldots ,v_l)$ be a generating family of the image of this orthogonal projection.

Let $Gr\in \mathbb{M}_{l}(\mathbb{C})$ be the 
Gram matrix associated with this basis,
namely
$Gr_{ij}=\langle v_i, v_j\rangle $,
 $X=(x_i)$ be the column vector given by 
$x_i=\langle v, v_i\rangle $, and
 $W$ be a symmetric real matrix satisfying $Gr \cdot W \cdot Gr = Gr$.

There exists a vector 
 $Y=(y_i)$ be a column vector such that $P(v)=\sum y_i v_i$. Besides this vector 
 can be chosen to satisfy
$$Y=W \cdot X.$$
\end{prop}

Note that we do not require the family $\{v_i\}$ to be independent in the hypothesis.

\begin{proof}
Although we will eventually need our proof in an Hermitian setting, let us prove it
in the Euclidean setting in order to simplify the notation. 
The proof in the Hermitian setting goes along the same lines
with the obvious modifications.

First, the existence of $Y$ is clear, as it is equivalent to the existence of an orthogonal
projection.
Now, let us observe that
$$X= Gr\cdot Y.$$
Indeed this is the matrix notation for the fact that $P(v)$ is characterized by
$$<P(v),v_i>=<v,v_i>$$ for all $i$, and this characterizes $P(v)$ uniquely together
with the fact that it is in $span \{v_i\}$.
Therefore $X\in \mathrm{Im} (Gr)$.
Now,  setting  $Y$ as $Y:=W\cdot X$ makes it an acceptable choice for $Y$,
in the sense that $X= Gr\cdot Y$. 
Indeed, this follows from the fact that
$Gr \cdot W \cdot Gr = Gr$
and that $X$ is in the image of $Gr$.
So this completes the proof.
\end{proof}

Now we are ready to proceed to the proof of Theorem \ref{Thm:CS}:

\begin{proof}[Proof of Theorem \ref{Thm:CS}:]

By the definition of $P$ in Lemma
\ref{lem-intermediate},
the canonical matrix element 
$P_{i_1,\ldots ,i_{2n};j_1,\ldots ,j_{2n}}$ is exactly 
$\int_{g \in O(d)} g_{i_1 j_1} \cdots g_{i_{2n} j_{2n}} d g$.
Equivalently stated, 
$$\int_{g \in O(d)} g_{i_1 j_1} \cdots g_{i_{2n} j_{2n}} d g=
<P(e_{i_1}\otimes\ldots \otimes e_{i_{2n}}),
e_{j_1}\otimes\ldots \otimes e_{j_{2n}}>$$ 
Let us first compute $P(e_{i_1}\otimes\ldots \otimes e_{i_{2n}})$.
According to Lemma \ref{prop:projection}, 
we have
$$\sum_{\mf{m}, \mf{n}\in \mcal{M}(2n)}
\rho (\mf{m})\cdot <\rho (\mf{n}),e_{i_1}\otimes\ldots \otimes e_{i_{2n}}>
\mr{Wg}^{O(d)}(\mf{m},\mf{n}).
$$
Taking the scalar product of the above vector equality 
against the vector $e_{j_1}\otimes\ldots \otimes e_{j_{2n}}$
gives 
$$
\int_{g \in O(d)} g_{i_1 j_1} \cdots g_{i_{2n} j_{2n}} d g =
\sum_{\mf{m}, \mf{n} \in \mcal{M}(2n)} 
\mr{Wg}^{O(d)}(\mf{m},\mf{n})
\prod_{k=1}^n \delta_{i_{\mf{m}(2k-1)}, i_{\mf{m}(2k)}} \delta_{j_{\mf{n}(2k-1)}, j_{\mf{n}(2k)}},
$$
as we can observe that  for
$\mf{m}$ in $\mcal{M}(2n)$, we have
 $$<e_{i_1}\otimes\ldots \otimes e_{i_{2n}},\rho (\mf{m})>=
 \prod_{k=1}^n \delta_{i_{\mf{m}(2k-1)}, i_{\mf{m}(2k)}} $$

\end{proof}

We would like to draw the attention of the reader of the fact that for the formula to holds,
it is not necessary to assume that 
$\mr{Wg}^{O(d)}_n$ is the pseudo-inverse of $G^{O(d)}_n$.
Actually, $\mr{Wg}^{O(d)}_n$ could be replaced by any matrix 
satisfying $G^{O(d)}_n W G^{O(d)}_n=G^{O(d)}_n$
without modifying the integration formula.
However, the pseudo-inverse is a canonical choice, and it will turn out to 
be a convenient on in the sequel of the paper.

Let us finish this section by recalling the Weingarten integration formula in the
unitary case. 

Let $\mcal{M}(2n)^U$ be the set of pair partitions on $\{1,2,\dots,2n\}$
pairing an element in $\{1,\ldots ,n\}$ with an element in 
$\{n+1,\ldots ,2n\}$. This is a subset of $\mcal{M}(2n)$.
For two pair partitions in $\mcal{M}(2n)^U$ as in the case
of $\mcal{M}(2n)$.
Similarly, we define the unitary gram matrix 
 $G^{U(d)}_n=(G^{O(d)}(\mf{m},\mf{n}))_{\mf{m},\mf{n} \in \mcal{M}(2n)^U}$ 
 by
$$
G^{U(d)}(\mf{m},\mf{n})= d^{\mr{loop}(\mf{m},\mf{n})},
$$
and the unitary Weingarten matrix  $\mr{Wg}^{U(d)}_n$ as 
the pseudo-inverse of $G^{U(d)}_n$.
The following Theorem describes the unitary integral case, and it
was proved in \cite{CollinsSniady}:

\begin{thm}[\cite{CollinsSniady}] \label{Thm:CS-Unitary}
Given indices $i_1,\dots,i_{2n}$ and $j_1,\dots,j_{2n}$ in 
$\{1,2,\dots,d\}$,
\begin{equation}
\int_{g \in U(d)} g_{i_1 j_1} \cdots g_{i_n j_n}
\overline{g_{i_{n+1} j_{n+1}}} \cdots
\overline{g_{i_{2n} j_{2n}}} d g =
\sum_{\mf{m}, \mf{n} \in \mcal{M}(2n)^U} 
\mr{Wg}^{U(d)}(\mf{m},\mf{n})
\prod_{k=1}^n \delta_{i_{\mf{m}(2k-1)}, i_{\mf{m}(2k)}} \delta_{j_{\mf{n}(2k-1)}, j_{\mf{n}(2k)}}. 
\end{equation}
\end{thm}

Note that this statement is very similar to the orthogonal case. 
The main difference is that we have to consider a submatrix of the original 
Gram matrix $G^{O(d)}$.
A proof similar to the orthogonal case could be carried along the same lines 
after adapting the invariant theory from the orthogonal case to the unitary case.

\section{A new expression for the orthogonal Weingarten function}
\label{sec:new-expression}

To each partition $\lambda=(\lambda_1,\lambda_2,\dots)$, 
we call  $\ell(\lambda)$ its {\it length} and $|\lambda|$ its {\it weight}.
If $|\lambda|=n$, we write $\lambda \vdash n$.
Let $H_n$ be the centralizer of the element $(1, 2) (3, 4 ) \cdots (2n-1, 2n)$ 
in $S_{2n}$,
where $(i,j)$ is a transposition. 
The group $H_n$ is called the {\it hyperoctahedral group}.
It is of order $2^n n!$
and isomorphic to the wreath product $S_2 \wr S_n$.
The pair $(S_{2n},H_n)$ is a {\it Gelfand pair}. 
Namely, the induced representation $\mr{ind}_{H_n}^{S_{2n}}(1_{H_n})$ of
the trivial representation $1_{H_n}$ of $H_n$ is a direct sum of inequivalent irreducible
representations of $S_{2n}$.

Following \cite[VII.1 and 2]{Mac},
let us review a few elements of the theory of the Gelfand pair $(S_{2n},H_n)$.
If each pair partition $\mf{m}$ in $\mcal{M}(2n)$ is expressed by the form
\eqref{eq:ExpressionPairPartition}
with $\mf{m}(2i-1) < \mf{m}(2i)$ for $1 \le i \le n$ and 
with $\mf{m}(1) <\mf{m}(3) < \cdots < \mf{m}(2n-1)$,
then we embed the set $\mcal{M}(2n)$ into $S_{2n}$ via the mapping
\begin{equation} \label{eq:identificationPM}
\mcal{M}(2n) \ni \mf{m} \mapsto
\begin{pmatrix}
1 & 2 & 3 & 4 & \cdots & 2n \\
\mf{m}(1) & \mf{m}(2) & \mf{m}(3) & \mf{m}(4) & \cdots & \mf{m}(2n) 
\end{pmatrix} \in S_{2n}.
\end{equation}
Pair partitions $\mf{m} \in \mcal{M}(2n)$ are representatives of the left cosets $\sigma H_n$
of $H_n$ in $S_{2n}$: 
\begin{equation} \label{eq:leftcosets}
S_{2n} = \bigsqcup_{\mf{m} \in \mcal{M}(2n)} \mf{m} H_n.
\end{equation}

Consider the double cosets $H_n \sigma H_n$ in $S_{2n}$.
These cosets are indexed by partitions of $n$:
\begin{equation}
S_{2n} = \bigsqcup_{\rho \vdash n} H_{\rho}. 
\end{equation}
Each double coset $H_\rho$ is of cardinality $|H_{\rho}|= \frac{(2^n n!)^2}{z_{2\rho}}$.
Here, for a partition $\mu$,  
$$
z_\mu=\prod_{i \ge 1} i^{m_i(\mu)} m_i(\mu)!
$$ 
with the multiplicity $m_i(\mu)$ of $i$ in $\mu$.
The permutation $\sigma \in S_{2n}$ is said to be of {\it coset-type} $\rho$
and written as $\Xi(\sigma)= \rho$
if $\sigma \in H_\rho$.

The partition $\Xi(\sigma)$ is also defined as follows.
Consider the graph $\Gamma(\sigma)$ whose vertex set is $\{1,2,\dots,2n\}$ and whose
edge set consists of $\{2i-1,2i\}$ and $\{\sigma(2i-1),\sigma(2i)\}$, $1 \le i \le n$.
Then $\Gamma(\sigma)$ has some connected components of even lengths 
$2 \rho_1 \ge 2 \rho_2 \ge \cdots$, say.
Thus $\sigma$ determines a partition $\rho=(\rho_1,\rho_2,\dots)$ of $n$.
This $\rho$ is the coset-type $\Xi(\sigma)$.
Two permutations $\sigma_1,\sigma_2 \in S_{2n}$ have the same coset-type if
and only if $\sigma_1 \in H_n \sigma_2 H_n$.

For each partition $\lambda$ of $n$, we define the {\it zonal spherical functions} of the 
Gelfand pair $(S_{2n},H_n)$ by
\begin{equation}
\omega^\lambda (\sigma) = \frac{1}{2^n n!} \sum_{\zeta \in H_n} \chi^{2\lambda}(\sigma \zeta),
\qquad \sigma \in S_{2n},
\end{equation}
where $\chi^{2\lambda}$ is the irreducible character of $S_{2n}$
associated with $2\lambda=(2\lambda_1,2\lambda_2,\dots)$.
These functions are constant on each double coset $H_\rho$.
We denote by $\omega^\lambda_\rho$ the value of $\omega^\lambda$ on the double coset $H_\rho$.
Put $f^{2\lambda}=\chi^{2 \lambda}(\mr{id}_n)$, where $\mr{id}_n$ is the identity permutation
in $S_n$.

For each $n \ge 1$ and each partition $\lambda \vdash n$, 
the {\it zonal polynomial} $Z_\lambda$ is a symmetric polynomial 
defined by
\begin{equation} \label{eq:Z}
Z_\lambda =2^n n! \sum_{\rho \vdash n }z_{2\rho}^{-1} \omega^\lambda_\rho p_{\rho},
\end{equation}
where $p_k(x_1,x_2,\dots)=x_1^k+x_2^k+\cdots$ is  the power-sum symmetric polynomial 
and $p_\rho=p_{\rho_1} p_{\rho_2} \cdots $.
The zonal polynomial $Z_\lambda$ is the Jack symmetric polynomial 
$J_{\lambda}^{(\alpha)}$ with parameter $\alpha=2$.
The specialization $Z_{\lambda}(1^d)=Z_{\lambda}(1,\dots,1)$ is given by
\begin{equation}
Z_{\lambda}(1^d)= \prod_{(i,j) \in \lambda} (d+2j-i-1),
\end{equation}
where $(i,j)$ run over all squares of the Young diagram of $\lambda$,
i.e., $i=1,2,\dots,\ell(\lambda)$ and $j=1,2,\dots,\lambda_i$.
In particular, $Z_{\lambda}(1^d)$ is zero if and only if $\ell(\lambda) >d$.

Now, entries of the Gram matrix are expressed by 
$G^{O(d)}(\mf{m},\mf{n})= d^{\ell(\Xi(\mf{m}^{-1}\mf{n}))}$. 
Here $\mf{m},\mf{n}$ are regarded as permutations in $S_{2n}$.
The following Theorem is our main result.

\begin{thm} \label{thm:orthogonalWg}
Let $d,n$ be positive integers.
For $\mf{m}, \mf{n} \in \mcal{M}(2n)$, 
we have
\begin{equation} \label{eq:Wgo}
\mr{Wg}^{O(d)}(\mf{m},\mf{n})= \frac{2^n n!}{(2n)!} 
\sum_{\begin{subarray}{c} \lambda \vdash n \\ \ell(\lambda) \le d 
\end{subarray}} 
f^{2\lambda} 
\frac{\omega^{\lambda}(\mf{m}^{-1} \mf{n})}{Z_\lambda(1^d)}.
\end{equation}
\end{thm}
Before proving this Theorem, in the forthcoming section
\ref{sec:proof}, let us illustrate it by an example:

\begin{example}
Let $n=2$. 
We have $\omega^{(2)}(\mf{m}^{-1} \mf{n})=1$ for any $\mf{m},\mf{n} \in \mcal{M}(4)$, and
$\omega^{(1,1)}(\mf{m}^{-1} \mf{n})=1$ if $\mf{m}=\mf{n}$, and 
$\omega^{(1,1)}(\mf{m}^{-1} \mf{n})=-\frac{1}{2}$ if $\mf{m} \not=\mf{n}$.
Therefore if $\mf{m}=\mf{n}$ and $d \ge 2$, we have
$$
\mr{Wg}^{O(d)} (\mf{m},\mf{n})= \frac{1}{3} \( \frac{1}{d(d+2)} + \frac{2}{d(d-1)}\)=
\frac{d+1}{d(d+2)(d-1)}. 
$$
If $\mf{m} \not=\mf{n}$ and $d \ge 2$, we have
$$
\mr{Wg}^{O(d)} (\mf{m},\mf{n})= \frac{1}{3} \( \frac{1}{d(d+2)} + 
\frac{2 \cdot (-\frac{1}{2})}{d(d-1)}\)=
\frac{-1}{d(d+2)(d-1)}. 
$$
If $d=1$, then $\mr{Wg}^{O(1)}(\mf{m},\mf{n})=\frac{1}{9}$ for any $\mf{m},\mf{n} \in \mcal{M}(4)$.
In general, for any $\mf{m}, \mf{n} \in \mcal{M}(2n)$ and $d=1$, we have
$$
\mr{Wg}^{O(1)} (\mf{m},\mf{n})= \frac{2^n n!}{(2n)!} f^{(2n)} 
\frac{\omega^{(n)}(\mf{m}^{-1} \mf{n})}{Z_{(n)}(1^1)} = \(\frac{2^n n!}{(2n)!} \)^2.
$$
\end{example}

To finish this section, observe that a result similar to 
Theorem \ref{thm:orthogonalWg} has already been obtained in  \cite{CollinsSniady}.
We identify the set $ \mcal{M}(2n)^U$ with the permutations on $n$ points
by associating to $\mf{m}\in  \mcal{M}(2n)^U$ the permutation $\sigma$
satisfying $\sigma (i)=j$ iff $\mf{m}$ links $i$ and $n+j$.
According to this identification, we can multiply 
elements of $ \mcal{M}(2n)^U$ according to the symmetric group structure, 
or consider their inverse. We can also consider their conjugacy classes.
With the above notation and notations parallel to our main Theorem
\ref{thm:orthogonalWg}, we have
\begin{thm}
\label{thm:unitaryWg}
Let $d,n$ be positive integers.
For $\mf{m}, \mf{n} \in \mcal{M}(2n)^U$, 
we have
\begin{equation} \label{eq:WgU}
\mr{Wg}^{U(d)}(\mf{m},\mf{n})= \frac{1}{n!^2} 
\sum_{\begin{subarray}{c} \lambda \vdash n \\ \ell(\lambda) \le d 
\end{subarray}} 
\frac{(f^\lambda)^2 \chi^{\lambda}(\mf{m}^{-1} \mf{n})}{s_{\lambda}(1^d)},
\end{equation}
where $s_{\lambda}$ is the Schur polynomial and $\chi^{\lambda}$
is the non-normalized character of the symmetric group.
\end{thm}
Although the proof of the above Theorem \ref{thm:unitaryWg}
was not so involved, the proof of the main Theorem 
\ref{thm:orthogonalWg} requires less standard algebraic material
such as Jack polynomial theory
(and in particular, zonal polynomials),  and Gelfand pair theory.
The next section is devoted to the review of this material and 
 the proof of Theorem \ref{thm:orthogonalWg}.

\section{Proof of Theorem \ref{thm:orthogonalWg}}
\label{sec:proof}

Our strategy of proof consists in representing $G_n^{O(d)}$ as an $H_n$ bi-invariant function on 
the symmetric group $S_{2n}$, and performing a Fourier transform on it.
Consider the $\bC$-vector space 
$$
\bC[\mcal{M}(2n)] = \bigoplus_{\mf{m} \in \mcal{M}(2n)} \bC \mf{m}
$$
and the representation $\rho_{2n}$ of $S_{2n}$ on $\bC[\mcal{M}(2n)]$ defined by
\begin{equation} \label{eq:reprho}
\rho_{2n} (\sigma) \mf{m} 
= \bigl\{ \{\sigma (\mf{m}(1)), \sigma (\mf{m}(2)) \}, \dots,
\{\sigma (\mf{m}(2n-1)), \sigma (\mf{m}(2n)) \} \bigr\}. 
\end{equation}
The pair partition $\mf{m}$ and $\rho_{2n}(\sigma)  \mf{m}$ are identified with 
elements of $S_{2n}$ via \eqref{eq:identificationPM}.
However, the permutation associated with $\rho_{2n}(\sigma) \mf{m}$ may
be different from the product $\sigma \mf{m}$ in $S_{2n}$.
By the left coset decomposition \eqref{eq:leftcosets}, 
there exists a unique permutation $\zeta_{\sigma,\mf{m}} \in H_n$ such that
\begin{equation}
\rho_{2n}(\sigma) \mf{m} = \sigma \mf{m} \zeta_{\sigma, \mf{m}} \in S_{2n}.
\end{equation}

\begin{example}
Let $\sigma =(1, 3, 2)(4)\in S_4$ and $\mf{m}= \{\{1,3\}, \{2,4\}\} \in \mcal{M}(4)$.
Then we have
$$
\rho_{4}(\sigma) \mf{m} = \{\{\sigma(1),\sigma(3)\}, \{\sigma(2),\sigma(4)\}\}=
\{\{3,2\}, \{1,4\}\} = \{\{1,4\}, \{2,3\}\},
$$
and so $\rho_4(\sigma) \mf{m}$ is 
$\( \begin{smallmatrix} 1 & 2 & 3 & 4 \\ 1 & 4 & 2 & 3 \end{smallmatrix}\) =(2 , 4 , 3)$ 
as an element in $S_4$,
which equals
$\sigma \mf{m} \zeta_{\sigma,\mf{m}}$ with 
$\zeta_{\sigma,\mf{m}}=( 1, 2) (1,3)(2, 4) \in H_2$.
\end{example}

Any matrix $L=(L(\mf{m},\mf{n}))_{\mf{m}, \mf{n} \in \mcal{M}(2n)}$ 
is identified with the endomorphism $T_L$ on $\bC[\mcal{M}(2n)]$
defined by
\begin{equation} \label{eq:defTL}
T_L ( \mf{m}) = \sum_{\mf{n} \in \mcal{M}(2n)} L(\mf{m},\mf{n}) \mf{n}.
\end{equation}
>From now,
we abbreviate the Gram matrix
$G^{O(d)}_n$ and Weingarten matrix $\mr{Wg}^{O(d)}_n$ to $G$ and $\mr{Wg}$, respectively.

\begin{lem} \label{lem:TGinvariant}
For the Gram matrix $G$,
we have
$T_{G}( \rho_{2n}(\sigma)\mf{m}) = \rho_{2n}(\sigma) T_{G}(\mf{m})$
for all $\sigma \in S_{2n}$ and $\mf{m} \in \mcal{M}(2n)$.
Hence $T_{G} \in \mr{End}_{\rho_{2n}(\bC[S_{2n}])} (\bC[\mcal{M}(2n)])$.
\end{lem}

\begin{proof}
For each $\sigma \in S_{2n}$ and $\mf{m}, \mf{n} \in \mcal{M}(2n)$,
we have
$$
G(\rho_{2n}(\sigma) \mf{m}, \rho_{2n}(\sigma) \mf{n}) 
= d^{\ell(\Xi( (\sigma \mf{m} \zeta_{\sigma,\mf{m}})^{-1} (\sigma \mf{n} \zeta_{\sigma,\mf{n}})))} 
=d^{\ell(\Xi(\zeta_{\sigma,\mf{m}}^{-1} \mf{m}^{-1} \mf{n} \zeta_{\sigma, \mf{n}}))}
=d^{\ell(\Xi(\mf{m}^{-1} \mf{n} ))} =G(\mf{m},\mf{n})
$$
because $\Xi(\zeta \sigma \zeta')=\Xi(\sigma)$ for all $\sigma \in S_{2n}$ and 
$\zeta, \zeta' \in H_n$.
Hence 
\begin{align*}
& T_G(\rho_{2n} (\sigma)\mf{m}) 
= \sum_{\mf{n} \in \mcal{M}(2n)} G(\rho_{2n}(\sigma) \mf{m}, \mf{n})\mf{n}
= \sum_{\mf{n} \in \mcal{M}(2n)} G(\rho_{2n}(\sigma) \mf{m}, \rho_{2n}(\sigma)\mf{n})
\rho_{2n}(\sigma) \mf{n} \\
=& \sum_{\mf{n} \in \mcal{M}(2n)} G(\mf{m}, \mf{n})\rho_{2n} (\sigma) \mf{n}
= \rho_{2n}(\sigma) T_G(\mf{m}).
\end{align*}
\end{proof}

Let
$$
e=\frac{1}{2^n n!} \sum_{\zeta \in H_n} \zeta.
$$
This element satisfies $e^2 =e$.
The set $\{\mf{m}e \ | \ \mf{m} \in \mcal{M}(2n)\}$ is a basis of
the left $\bC[S_{2n}]$-module $\bC[S_{2n}]e$.
The proof of the following lemma can be found, for example, in
\cite[VII.1]{Mac}:

\begin{lem}  \label{lem:Bi-invariant}
The endomorphism ring $\mr{End}_{\bC[S_{2n}]} (\bC[S_{2n}]e)$
is anti-isomorphic to $e \bC[S_{2n}]e$,
the anti-isomorphism being $\phi \mapsto \phi(e)$.
\end{lem}

By definition, zonal spherical functions $\omega^\lambda$ for the Gelfand pair $(S_{2n},H_n)$
are expressed as 
$\omega^\lambda= \chi^{2\lambda} e$.

\begin{lem} \label{lem:zonal}
Zonal spherical functions $\omega^\lambda$ satisfy the
following properties.
\begin{enumerate}
\item $\omega^\lambda \omega^\mu = \delta_{\lambda \mu} \frac{(2n)!}{f^{2\lambda}} \omega^\lambda$.
\item $\{\omega^\lambda \ | \ \lambda \vdash n\}$ is a basis of  
$e \bC[S_{2n}]e$.
\item Let $\mu=(\mu_1,\mu_2,\dots) \vdash n$, and 
let $p_\mu=p_{\mu_1} p_{\mu_2}\cdots p_{\ell(\mu)}$ be the power-sum symmetric polynomial.
Then we have
\begin{equation} \label{eq:PowerZonal}
p_\mu(x_1,x_2,\dots,x_r) = \frac{2^n n!}{(2n)!} 
\sum_{\lambda \vdash n} f^{2\lambda}
\omega^\lambda_\mu Z_\lambda(x_1,\dots,x_r).
\end{equation}
Here $x_1,\dots,x_r$ are indeterminates and $r$ is a positive integer.
\end{enumerate}
\end{lem}
\begin{proof}
The first and second claims are seen in \cite[VII (1.4)]{Mac}.
The third claim is seen in \cite[VII (2.16)]{Mac}.
\end{proof}

Let $\mcal{I}$ be the bijective linear map from $\bC[\mcal{M}(2n)]$ to $\bC[S_{2n}]e$
defined by
$\mcal{I}(\mf{m}) = \mf{m} e \ (\mf{m} \in \mcal{M}(2n))$.

\begin{lem} \label{lem:intertw}
The map $\mcal{I}$ is an intertwiner.
Hence $\bC[\mcal{M}(2n)]$ with the action $\rho_{2n}$
is isomorphic to $\bC[S_{2n}]e$  
as $\bC[S_{2n}]$-modules. 
\end{lem}

\begin{proof}
For each $\sigma \in S_{2n}$ and $\mf{m} \in \mcal{M}(2n)$,
we have
$$
\mcal{I}(\rho_{2n}(\sigma)\mf{m})= (\rho(\sigma)\mf{m})e
= (\sigma \mf{m} \zeta_{\sigma,\mf{m}}) e = 
\sigma (\mf{m} 
\zeta_{\sigma,\mf{m}} e)
=\sigma (\mf{m} e)  = \sigma \mcal{I}(\mf{m})   
$$
since $\zeta e =e$ for all $\zeta \in H_n$.
\end{proof}

By Lemma \ref{lem:TGinvariant} and Lemma \ref{lem:intertw}, the endomorphism 
$\mcal{I} \circ T_G \circ \mcal{I}^{-1}$ belongs to 
$\mr{End}_{\bC[S_{2n}]} (\bC[S_{2n}]e)$.
Moreover, it follows from Lemma \ref{lem:Bi-invariant} that
$\hat{G} :=(\mcal{I} \circ T_G \circ \mcal{I}^{-1})(e)$ belongs to  
$e \bC[S_{2n}] e$ and 
$$
\hat{G} =\sum_{\mf{m} \in \mcal{M}(2n)} d^{\ell( \Xi(\mf{m}))} \mf{m} e.
$$
If we specialize as $(x_1,x_2,\dots)=(1^d)$ in \eqref{eq:PowerZonal}, we obtain 
\begin{equation} \label{eq:expand_d_Z}
d^{\ell(\mu)} = \frac{2^n n!}{(2n)!} 
\sum_{\begin{subarray}{c} \lambda \vdash n \\ \ell(\lambda) \le d \end{subarray}} 
f^{2\lambda}
\omega^\lambda_\mu Z_\lambda(1^d).
\end{equation}
Therefore we obtain 
$$
\hat{G} = 
  \frac{2^n n!}{(2n)!} 
\sum_{\begin{subarray}{c} \lambda \vdash n \\ \ell(\lambda) \le d \end{subarray}} 
 f^{2\lambda} 
Z_\lambda(1^d) \sum_{\mf{m} \in \mcal{M}(2n)} \omega^\lambda_{\Xi(\mf{m})} 
\mf{m} e.
$$
Since, by the left coset decomposition \eqref{eq:leftcosets},
$$
2^n n! \sum_{\mf{m} \in \mcal{M}(2n)} \omega^\lambda_{\Xi(\mf{m})} 
\mf{m} e 
= \sum_{\mf{m} \in \mcal{M}(2n)} \sum_{\zeta \in H_n} \omega^\lambda( \mf{m}\zeta) \mf{m} \zeta
= \sum_{\sigma \in S_{2n}} \omega^\lambda(\sigma) \sigma=
\omega^\lambda,
$$
we can express as
\begin{equation}
\hat{G}= 
\frac{1}{(2n)!} 
\sum_{\begin{subarray}{c} \lambda \vdash n \\ \ell(\lambda) \le d \end{subarray}} f^{2\lambda} 
Z_\lambda(1^d) \omega^\lambda. 
\end{equation}

Define the element $\hat{\mathrm{Wg}}$ in $e \bC[S_{2n}] e$ by
\begin{equation}
\hat{\mathrm{Wg}} = \frac{1}{(2n)!}
\sum_{\begin{subarray}{c} \lambda \vdash n \\ \ell(\lambda) \le d \end{subarray}} f^{2\lambda} 
Z_\lambda(1^d)^{-1} \omega^\lambda. 
\end{equation}
By Claim 1 of Lemma \ref{lem:zonal} we obtain the following proposition.

\begin{prop} \label{lem:eAe}
It holds that 
$$
\hat{G} \cdot \hat{\mathrm{Wg}}= \frac{1}{(2n)!}
\sum_{\begin{subarray}{c} \lambda \vdash n \\ \ell(\lambda) \le d \end{subarray}}
f^{2\lambda} \omega^\lambda.
$$ 
Moreover, we have 
$$
\hat{G} \cdot \hat{\mathrm{Wg}} \cdot \hat{G}= \hat{G}, \qquad 
\hat{\mathrm{Wg}} \cdot \hat{G} \cdot \hat{\mathrm{Wg}}=\hat{\mathrm{Wg}}.
$$
\end{prop}

As the matrix $G$ is corresponding to $\hat{G}$, 
the matrix $\mr{Wg}$ given in \eqref{eq:Wgo} is corresponding to $\hat{\mr{Wg}}$.
Hence, since $G$ is symmetric, 
$\mr{Wg}$ is the pseudo-inverse matrix of $G$ by Proposition \ref{lem:eAe}.

Using Claim 1 and 2 of Lemma \ref{lem:zonal},
we obtain 
\begin{equation}
e=\frac{1}{(2n)!} \sum_{\lambda \vdash n} f^{2\lambda} \omega^\lambda
\end{equation}
because $e^2=e$. 
The element $e$ is the identity of the algebra $e \bC[S_{2n}]e$
and $\hat{G} \cdot \hat{\mr{Wg}}=e$ if $d \ge n$.
Hence, when $d \ge n$, the matrix $G$ is invertible and $\mr{Wg}=G^{-1}$.

\section{Moments of the trace of a truncated orthogonal matrix}
\label{sec:truncated}

As an application of Theorem \ref{Thm:CS} and \ref{thm:orthogonalWg},
we calculate the moment of the trace of a truncated orthogonal random matrix.

\begin{thm} \label{thm:truncated}
Let $k,d,n$  be positive integers and suppose $k \le d$. 
Let $g=(g_{ij})_{1 \le i,j \le d}$ 
be a random matrix chosen with respect to the Haar measure from the orthogonal group $O(d)$,
and let $g^{(k)}$ be its $k \times k$ upper left corner: $g^{(k)}=(g_{ij})_{1 \le i,j \le k}$.
Then we have
\begin{equation}
\int_{g \in O(d)}  (\tr (g^{(k)}))^{2n} d g = 
\sum_{\begin{subarray}{c} \lambda \vdash n \\ \ell(\lambda) \le k \end{subarray}}
f^{2\lambda} \frac{Z_{\lambda}(1^k)}{Z_{\lambda}(1^d)}.
\end{equation}
\end{thm}

We recover Rains's result in \cite{Rains} by setting $k=d$ in Theorem \ref{thm:truncated}.
The average of the odd power of $\tr(g^{(k)})$ is zero:
\begin{equation}
\int_{g \in O(d)}  (\tr (g^{(k)}))^{2n-1} d g = 0.
\end{equation}
The unitary group version for Theorem \ref{thm:truncated}
is seen in \cite[Theorem 1]{Novak}.

\begin{proof}
A straightforward calculation and Theorem \ref{Thm:CS} give
\begin{align}
&  \int_{g \in O(d)} (\tr (g^{(k)}))^{2n} d g 
= \sum_{a_1=1}^k \cdots \sum_{a_{2n}=1}^k \int_{g \in O(d)} 
g_{a_1 a_1} \cdots g_{a_{2n} a_{2n}} d g  \notag \\
=& \sum_{a_1=1}^k \cdots \sum_{a_{2n}=1}^k \sum_{\mf{m},\mf{n} \in \mcal{M}(2n)} 
\mr{Wg}^{O(d)}(\mf{m},\mf{n}) \prod_{p=1}^{n} \delta_{a_{\mf{m}(2p-1)},a_{\mf{m}(2p)}}
\delta_{a_{\mf{n}(2p-1)},a_{\mf{n}(2p)}} \notag \\
=& \sum_{\mf{m},\mf{n} \in \mcal{M}(2n)} 
\mr{Wg}^{O(d)}(\mf{m},\mf{n}) \sum_{a_1=1}^k \cdots \sum_{a_{2n}=1}^k 
\prod_{p=1}^{n} \delta_{a_{\mf{m}(2p-1)},a_{\mf{m}(2p)}}
\delta_{a_{\mf{n}(2p-1)},a_{\mf{n}(2p)}} \notag \\
=& \sum_{\mf{m},\mf{n} \in \mcal{M}(2n)} 
\mr{Wg}^{O(d)}(\mf{m},\mf{n}) k^{\mr{loop}(\mf{m},\mf{n})}. \label{eq:Tr_G_Wg}
\end{align}
This coincides with the trace of the matrix $\mr{Wg}^{O(d)}_n \cdot G^{O(k)}_n$.

By \eqref{eq:expand_d_Z}, we have
$$
k^{\mr{loop}(\mf{m},\mf{n})} = k^{\ell(\Xi (\mf{m}^{-1} \mf{n}))} = 
\frac{2^n n!}{(2n)!} 
\sum_{\begin{subarray}{c} \mu \vdash n \\ \ell(\mu) \le k \end{subarray}} 
f^{2\mu}
Z_\mu(1^k) \omega^\mu(\mf{m}^{-1} \mf{n}).
$$
It follows from this equation and Theorem \ref{thm:orthogonalWg} that
\eqref{eq:Tr_G_Wg} equals
$$
\(\frac{2^n n!}{(2n)!}\)^2 
\sum_{\begin{subarray}{c} \lambda \vdash n \\ \ell(\lambda) \le d \end{subarray}} 
\sum_{\begin{subarray}{c} \mu \vdash n \\ \ell(\mu) \le k \end{subarray}}
f^{2\lambda} f^{2\mu} \frac{Z_\mu(1^k)}{Z_\lambda(1^d)} 
\sum_{\mf{m},\mf{n} \in \mcal{M}(2n)} 
\omega^{\lambda} (\mf{m}^{-1} \mf{n}) \omega^{\mu} (\mf{m}^{-1} \mf{n}).
$$
Now it is enough to prove the following identity:
for any $\lambda, \mu \vdash n$, 
\begin{equation} \label{eq:OrthogonalityOnM}
\sum_{\mf{m},\mf{n} \in \mcal{M}(2n)} 
\omega^{\lambda} (\mf{m}^{-1} \mf{n}) \omega^{\mu} (\mf{m}^{-1} \mf{n})
= \( \frac{(2n)!}{2^n n!}\)^2 \frac{1}{f^{2\lambda}} \delta_{\lambda \mu}.
\end{equation}

By the coset decomposition in \eqref{eq:leftcosets}, we see that 
\begin{align*}
&\sum_{\mf{m},\mf{n} \in \mcal{M}(2n)} 
\omega^{\lambda} (\mf{m}^{-1} \mf{n}) \omega^{\mu} (\mf{m}^{-1} \mf{n}) \\
=&\( \frac{1}{2^n n!} \)^2
\sum_{\zeta, \zeta' \in H_n} 
\sum_{\mf{m},\mf{n} \in \mcal{M}(2n)} 
\omega^{\lambda} ((\mf{m}\zeta)^{-1} (\mf{n}\zeta')) 
\omega^{\mu} ((\mf{m}\zeta)^{-1} (\mf{n}\zeta')) \\
=& \( \frac{1}{2^n n!} \)^2 \sum_{\sigma, \tau \in S_{2n}}
\omega^\lambda (\sigma^{-1} \tau) \omega^\mu (\sigma^{-1} \tau) \\
=& \frac{(2n)!}{(2^n n!)^2}  \sum_{\sigma\in S_{2n}}
\omega^\lambda (\sigma) \omega^\mu (\sigma) 
= \(\frac{(2n)!}{2^n n!} \)^2 \langle \omega^\lambda, \omega^\mu \rangle_{S_{2n}}.
\end{align*}
Here $\langle \cdot \, ,  \cdot \rangle_{S_{2n}}$ is the inner product on $\bC[S_{2n}]$. 
Since $\langle \omega^\lambda, \omega^\mu \rangle_{S_{2n}} = \frac{1}{f^{2\lambda}}
\delta_{\lambda \mu}$ (see \cite[VII (1.4) (iv) ]{Mac}),
the equality \eqref{eq:OrthogonalityOnM} holds true.
\end{proof}

Observe that in the first computation of the above proof also lead in the earlier 
paper \cite{BanicaCollins} to the notion of truncated characters for
compact quantum groups. 

\section{The full cycle case}
\label{sec:full-cycle}

In this section,
we suppose $d \ge n$.
We are interested in 
$\mr{Wg}^{O(d)}(\mf{m},\mf{n})$
where the coset type of $\mf{m}^{-1}\mf{n}$ is the $1$-length partition $(n)$. 
The motivation for this comes from the paper \cite{Collins}.
Indeed, in the unitary case, the following was proved:

\begin{prop}
Given a partition $\mu$ of $n$,
let $\mr{Wg}^{U}(\mu,d):= \mr{Wg}^{U(d)}(\mf{m},\mf{n})$,
where $\mf{m},\mf{n} \in \mcal{M}(2n)^U$ and the coset type of $\mf{m}^{-1}\mf{n}$ is $\mu$.
(Recall \eqref{eq:WgU}). 
Then the following holds true:
\begin{equation}
\mr{Wg}^{U}((n),d)=(-1)^{n-1}c_{n-1}\prod_{-n+1\leq j \leq n-1}(d-j)^{-1}
\end{equation}
Here $c_k=\frac{(2k)!}{(k+1)! \, k!}$ is the Catalan number.
\end{prop}

\begin{proof}[Sketch of proof]
Indeed, a classical combinatorial result about the Schur polynomials
shows that if one writes $\mr{Wg}^{U}((n),d)$ as an irreducible rational fraction,
its denominator has to be $\prod_{-q+1\leq j \leq q-1}(d-j)$. 
An explicit formula for the aysmptotic value as $d\to\infty$ of
$\mr{Wg}^{U}((n),d)$
given in \cite{Collins}, together with the fact that 
$\mr{Wg}^{U}((n),d)$ has to be a rational fraction concludes the proof.
\end{proof}

Note that we found it conceptually strange that an asymptotic estimate
was necessary to prove such an exact algebraic formula.
One can actually observe that the knowledge of asymptotics can actually
be bypassed by an application of the Zeilberger algorithms implemented
on many formal algebraic computing softwares (e.g. maple or mathematica).
See \cite{Zeilberger} for details.
Related generalizations of this results are in preparation
in a paper by the second author and Novak, \cite{MatsumotoNovak}.

In the orthogonal case, the entries $\mr{Wg}^{O(d)}(\mf{m},\mf{n})$ of the Weingarten matrix 
only depend on the coset type $\Xi(\mf{m}^{-1} \mf{n})$.
For each partition $\mu$ of $n$ and a positive integer $d \ge n$, we put
\begin{equation} 
\mr{Wg}^{O}(\mu,d)= \frac{1}{(2n-1)!!} \sum_{\lambda \vdash n} 
\frac{f^{2\lambda} \omega^\lambda_\mu}{Z_\lambda(1^d)}.
\end{equation}
Theorem \ref{thm:orthogonalWg} implies that if $\mf{m},\mf{n} \in \mcal{M}(2n)$ and 
$\mu=\Xi(\mf{m}^{-1}\mf{n})$ then
$\mr{Wg}^{O}(\mu,d)=\mr{Wg}^{O(d)}(\mf{m},\mf{n})$.

Consider $\mr{Wg}^{O}((n),d)$. We know values $\omega^{\lambda}_{(n)}$ 
(e.g. \cite[VII.2, Example 2(c)]{Mac}):
$$
\omega^\lambda_{(n)}=\frac{1}{2^{n-1} (n-1)!} 
\prod_{\begin{subarray}{c} (i,j) \in \lambda \\ (i,j) \not=(1,1) \end{subarray}}
(2j-i-1).
$$
In paticular, $\omega^{\lambda}_{(n)}=0$ if the square $(3,2)$ is contained 
in the Young diagram of $\lambda$,
i.e., $\lambda_3 \ge 2$.
Therefore we obtain an expression
\begin{equation} \label{eq:FullCycleWg}
\mr{Wg}^{O}((n),d)= \frac{1}{(2n-1)!} 
\sum_{\begin{subarray}{c} \lambda \vdash n \\ \lambda_3 \le 1 \end{subarray}}
\frac{f^{2\lambda}}{d} 
\prod_{\begin{subarray}{c} (i,j) \in \lambda \\ (i,j) \not=(1,1) \end{subarray}}
\frac{2j-i-1}{d+2j-i-1}.
\end{equation}
If we reduce to a common denominator, the denominator has the form
$\prod_{(i,j)}(d+2j-i-1)$, where $(i,j)$ run over the set
\begin{align*}
&\{(i,j) \ | \ \text{there exists $\lambda \vdash n$ such that $\lambda_3 \le 1$ and 
$(i,j) \in \lambda$} \} \\
=& \{(1,j) \ | \ j=1,2,\dots, n \} \sqcup \{(i,1) \ | \ i=2,3,\dots,n\} \sqcup 
\{(2,j) \ | \ j=2,3,\dots, \lfloor \frac{n}{2} \rfloor \}.
\end{align*}
Now we define the polynomial $P_n(d)$ in $d$ with $\bQ$-coefficients via 
\begin{equation} \label{eq:PolynomialP}
\mr{Wg}^{O}((n),d)= \frac{(-1)^{n-1} P_n(d)}
{d \prod_{j=1}^{n-1} (d+2j)(d-j) \cdot \prod_{k=1}^{\lfloor \frac{n}{2} \rfloor -1}
(d+2k-1)}.
\end{equation}

\begin{prop}
The polynomials $P_n(d)$ satisy the following properties.
\begin{itemize}
\item[(i)] For $n \ge 2$, the degree of $P_n$ is exactly $\lfloor \frac{n}{2} \rfloor -1$.
\item[(ii)] The coefficient of the highest-degree term of $P_n$ equals the Catalan number $c_{n-1}$.
\item[(iii)] The constant term of $P_n$ equals 
$$
P_n(0)= \sum_{\begin{subarray}{c} \lambda \vdash n \\ \lambda_3 \le 1 \end{subarray}}
f^{2\lambda} \times
\frac{(n-1)!}{(2n-1)!!} \times 
\begin{cases} 
(n-4)!! & \text{if $n$ is odd and $n \ge 5$} \\
(n-3)!! & \text{if $n$ is even and $n \ge 4$} \\
1 & \text{otherwise}.
\end{cases}
$$
\end{itemize}
\end{prop}

\begin{proof}
We use a result for the asymptotics of the Weingarten function, obtained in \cite{CollinsSniady}:
for a partition $\mu$ of $n$, as $d \to \infty$, 
$$
\mr{Wg}^{O}(\mu,d) = \( \prod_{i \ge 1} (-1)^{\mu_i-1} c_{\mu_i-1}\) d^{-2n+\ell(\mu)}
(1+\mr{O}(d^{-1})).
$$ 
In particular, $\mr{Wg}^{O}((n),d)=(-1)^{n-1} c_{n-1} d^{-2n+1}(1+\mr{O}(d^{-1}))$.
This fact and equation \eqref{eq:PolynomialP} imply our desired first and second claims.
The third claim follows from \eqref{eq:FullCycleWg} and \eqref{eq:PolynomialP} with $ d \to 0$.
\end{proof}

Suppose $\lambda \vdash n$ satisfies $\lambda_3 \le 1$.
Then we can write as $\lambda=(r,s,1^{n-r-s})$ with
$r=1,2,\dots,n$ and $s=0,1,\dots,\min\{r,n-r\}$.
By the well-known hook formula for $f^{2\lambda}$, we obtain that if $s \ge 1$,
\begin{align*}
& hook(r,s,n):= \frac{(2n)!}{f^{2\lambda}}=\frac{(2n)!}{f^{(2r,2s,2^{n-r-s})}} \\
=&
(n+r-s+1)(n+r-s)(n-r+s)(n-r+s-1)  \cdot (n-r-s+1)! \cdot (n-r-s)!\\
& \quad \cdot (2s-2)! \cdot
\frac{(2r-1)!}{2r-2s+1}.
\end{align*}
Now by \eqref{eq:FullCycleWg} and \eqref{eq:PolynomialP} we have
\begin{align}
P_n(d) =&(-1)^{n-1} 2n \prod_{j=1}^{n-1}(d+2j)(d-j) \cdot 
\prod_{k=1}^{\lfloor \frac{n}{2} \rfloor-1} (d+2k-1)  
 \Bigg(\frac{1}{(2n)!} \prod_{j=2}^{n} \frac{2j-2}{d+2j-2}  \\
& +\sum_{r=1}^{n-1} \sum_{s=1}^{\min (r,n-r)} \frac{1}{hook(r,s,n)} 
\prod_{j=2}^{r} \frac{2j-2}{d+2j-2} \cdot \prod_{j=1}^{s}\frac{2j-3}{d+2j-3} \cdot 
\prod_{i=3}^{n-r-s+2} \frac{-(i-1)}{d-(i-1)}\Bigg). \notag
\end{align}
For example,
\begin{align*}
P_1(d)=&P_2(d)= 1. \qquad
P_3(d)= 2. \qquad
P_4(d)= 5d+6. \qquad
P_5(d)= 14d+24. \\
P_6(d)=& 42d^2+236d+344. \qquad
P_7(d)= 132d^2+920d+1824. \\
P_8(d)=& 429d^3+5924d^2+29116d+51600. \\
P_9(d)=& 1430d^3+ 23124d^2+138352d +305280.\\
P_{10}(d)=& 4862d^4 +126816d^3+1326016 d^2 +6598896d+13071744.
\end{align*}
From those examples, we suggest the following conjecture.

\begin{conj}
All of coefficients of $P_n(d)$ are nonnegative integers.
\end{conj}

\section{Table of value}
\label{sec:tableofvalues}

We finish this paper by giving a table of values of the orthogonal Weingarten functions up to $n= 6$.
To do it, we employ the table of zonal polynomials \cite{PJ}.
Observe that even with $n=4$, this is a quite heavy formal computation without our formula,
and a standard formal algebra system on a modern desktop can not go beyond $n=4$
($105\times 105$ formal matrix to invert) without our formula provided in this paper.
We expect that our formula should have quite interesting applications quite soon.
The table for $n \le 4$ is also seen in \cite{CollinsSniady}.

Table goes here:
\begin{align*}
\mr{Wg}^{O}((1),d)=& \frac{1}{d}. \\
\mr{Wg}^{O}((2),d)=&  \frac{1}{3} \( \frac{1 \cdot 1}{d(d+2)} + 
\frac{2 \cdot \(-\frac{1}{2}\)}{d(d-1)} \)
= \frac{-1}{d(d+2)(d-1)}. \\
\mr{Wg}^{O}((1^2),d)=&  \frac{1}{3} \( \frac{1 \cdot 1}{d(d+2)} + 
\frac{2 \cdot 1}{d(d-1)} \)
= \frac{d+1}{d(d+2)(d-1)}. 
\end{align*}
\begin{align*}
\mr{Wg}^{O}((3),d)=&  \frac{1}{15} \( \frac{1 \cdot 1}{d(d+2)(d+4)} + 
\frac{9 \cdot \(-\frac{1}{4}\)}{d(d+1)(d-1)} + \frac{5 \cdot \frac{1}{4}}{d(d-1)(d-2)} \) \\
=& \frac{2}{d(d+2)(d+4)(d-1)(d-2)}.\\
\mr{Wg}^{O}((2,1),d)=&  \frac{1}{15} \( \frac{1 \cdot 1}{d(d+2)(d+4)} + 
\frac{9 \cdot \frac{1}{6}}{d(d+1)(d-1)} + \frac{5 \cdot \(-\frac{1}{2}\)}{d(d-1)(d-2)} \) \\
=& \frac{-1}{d(d+4)(d-1)(d-2)}. \\
\mr{Wg}^{O}((1^3),d)=&  \frac{1}{15} \( \frac{1 \cdot 1}{d(d+2)(d+4)} + 
\frac{9 \cdot 1}{d(d+1)(d-1)} + \frac{5 \cdot 1}{d(d-1)(d-2)} \) \\
=& \frac{d^2+3d-2}{d(d+2)(d+4)(d-1)(d-2)}. 
\end{align*}
\begin{align*}
\mr{Wg}^{O}((4),d)
=& \frac{-(5d+6)}{d(d+1)(d+2)(d+4)(d+6)(d-1)(d-2)(d-3)}. \\
\mr{Wg}^{O}((3,1),d)=& \frac{2}{(d+1)(d+2)(d+6)(d-1)(d-2)(d-3)}. \\
\mr{Wg}^{O}((2,2),d)=& \frac{d^2+5d+18}{d(d+1)(d+2)(d+4)(d+6)(d-1)(d-2)(d-3)}. \\
\mr{Wg}^{O}((2,1,1),d)=& \frac{-d^3-6d^2-3d+6}{d(d+1)(d+2)(d+4)(d+6)(d-1)(d-2)(d-3)}. \\
\mr{Wg}^{O}((1^4),d)
=& \frac{(d+3)(d^2+6d+1)}{d(d+1)(d+2)(d+4)(d+6)(d-1)(d-3)}.
\end{align*}

\begin{align*}
\mr{Wg}^{O}((5),d)=& \frac{2(7d+12)}{d(d+1)(d+2)(d+4)(d+6)(d+8)(d-1)(d-2)(d-3)(d-4)}. \\
\mr{Wg}^{O}((4,1),d)=& \frac{-5d+4}{d(d+1)(d+2)(d+4)(d+8)(d-1)(d-2)(d-3)(d-4)}. \\
\mr{Wg}^{O}((3,2),d)=& \frac{-2(d^2+7d+36)}{d(d+1)(d+2)(d+4)(d+6)(d+8)(d-1)(d-2)(d-3)(d-4)}. \\
\mr{Wg}^{O}((3,1^2),d)=& \frac{2(d^3+8d^2+d-36)}{d(d+1)(d+2)(d+4)(d+6)(d+8)(d-1)(d-2)(d-3)(d-4)}. \\
\mr{Wg}^{O}((2^2,1),d)=& \frac{d^2+3d+4}{d(d+1)(d+2)(d+4)(d+8)(d-1)(d-2)(d-3)(d-4)}. \\
\mr{Wg}^{O}((2,1^3),d)=& \frac{-d^4-10d^3-7d^2+86d+24}{d(d+1)(d+2)(d+4)(d+6)(d+8)(d-1)(d-2)(d-3)(d-4)}. \\
\mr{Wg}^{O}((1^5),d)=& \frac{d^5+11d^4+5d^3-175d^2-122d+408}{d(d+1)(d+2)(d+4)(d+6)(d+8)(d-1)(d-2)(d-3)(d-4)}.
\end{align*}

{\footnotesize
\begin{align*}
\mr{Wg}^{O}((6),d)=& \frac{-2(21d^2+118d+172)}
{d(d+1)(d+2)(d+3)(d+4)(d+6)(d+8)(d+10)(d-1)(d-2)(d-3)(d-4)(d-5)}.\\
\mr{Wg}^{O}((5,1),d)=& \frac{2(7d^3+12d^2-35d-10)}
{d^2(d+1)(d+2)(d+3)(d+4)(d+6)(d+10)(d-1)(d-2)(d-3)(d-4)(d-5)}.\\
\mr{Wg}^{O}((4,2),d)=& \frac{5d^4+61d^3+406d^2+840d+640}
{d^2(d+1)(d+2)(d+3)(d+4)(d+6)(d+8)(d+10)(d-1)(d-2)(d-3)(d-4)(d-5)}.\\
\mr{Wg}^{O}((4,1^2),d)=& \frac{-5d^5-66d^4-131d^3+642d^2+1272d-640}
{d^2(d+1)(d+2)(d+3)(d+4)(d+6)(d+8)(d+10)(d-1)(d-2)(d-3)(d-4)(d-5)}. \\
\mr{Wg}^{O}((3^2),d)=& \frac{4(d^4+13d^3+117d^2+300d-240)}
{d^2(d+1)(d+2)(d+3)(d+4)(d+6)(d+8)(d+10)(d-1)(d-2)(d-3)(d-4)(d-5)}.\\
\mr{Wg}^{O}((3,2,1),d)=& \frac{2(-d^4-7d^3-24d^2+12d+60)}
{d^2(d+1)(d+2)(d+3)(d+4)(d+6)(d+10)(d-1)(d-2)(d-3)(d-4)(d-5)}.\\
\mr{Wg}^{O}((3,1^3),d)=& \frac{2(d^6+16d^5+49d^4-200d^3-810d^2-96d+960)}
{d^2(d+1)(d+2)(d+3)(d+4)(d+6)(d+8)(d+10)(d-1)(d-2)(d-3)(d-4)(d-5)}.\\
\mr{Wg}^{O}((2^3),d)=& \frac{-(d^5+16d^4+101d^3+394d^2+2408d+3840)}
{d^2(d+1)(d+2)(d+3)(d+4)(d+6)(d+8)(d+10)(d-1)(d-2)(d-3)(d-4)(d-5)}.\\
\mr{Wg}^{O}((2^2,1^2),d)=& \frac{d^6+17d^5+77d^4+7d^3-446d^2-472d-1280}
{d^2(d+1)(d+2)(d+3)(d+4)(d+6)(d+8)(d+10)(d-1)(d-2)(d-3)(d-4)(d-5)}.\\
\mr{Wg}^{O}((2,1^4),d)=& \frac{-d^5-22d^4-154d^3-316d^2+339d+1146}
{d(d+1)(d+2)(d+3)(d+4)(d+6)(d+8)(d+10)(d-1)(d-2)(d-3)(d-5)}.\\
\mr{Wg}^{O}((1^6),d)=& \frac{d^8+19d^7+68d^6-490d^5-2687d^4+1807d^3+17754d^2 +6120d-15360}
{d^2(d+1)(d+2)(d+3)(d+4)(d+6)(d+8)(d+10)(d-1)(d-2)(d-3)(d-4)(d-5)}.
\end{align*}
}

\section*{Acknowledgements}
Both authors would like to thank Prof A. Hora for inviting them to give lectures at
Nagoya University in December 2007; this is how the project of this paper got started.
Later on, S.M. visited the university of Ottawa in August 2008 and B.C. visited a second time 
Nagoya University
in March 2009 in the framework of this collaboration. Both authors are grateful 
to the respective institutions for their hospitalities. 

The research of B.C. was partly supported by an
NSERC discovery grant RGPIN/341303-2007, and 
ANR projects GALOISINT and GRANMA.

The research of S.M. was partly supported by Grant-in-Aid for JSPS Fellows no. 20001840.


\noindent
\textsc{Beno\^{i}t Collins: 
D\'epartement de Math\'ematique et Statistique, Universit\'e d'Ottawa,
585 King Edward, Ottawa, ON, K1N6N5 Canada
and 
CNRS, Institut Camille Jordan Universit\'e  Lyon 1, 43 Bd du 11 Novembre 1918, 69622 Villeurbanne, 
France}
E-mail: \verb|bcollins@uottawa.ca| 

\medskip

\noindent
\textsc{Sho Matsumoto: 
Graduate School of Mathematics, Nagoya University, Nagoya, 464-8602, Japan.} 
E-mail: \verb|sho-matsumoto@math.nagoya-u.ac.jp| 

\end{document}